\documentclass{osa-article}

\journal{oe}

\articletype{Research Article}

\usepackage{url}

\begin{document}

\title{300~GHz wireless link based on an integrated Kerr soliton comb}

\author{Tomohiro Tetsumoto,\authormark{1} and Antoine Rolland\authormark{1*}}

\address{\authormark{1}IMRA America Inc., Boulder Research Labs, 1551 South Sunset St, Suite C, Longmont, Colorado 80501, USA}

\email{\authormark{*}arolland@imra.com} 



\begin{abstract} 
A Kerr microresonator frequency comb has enabled the generation of low-phase-noise millimeter- and terahertz-waves in conjunction with an ultrafast photodiode.
It is intriguing to employ the new light source in wireless communication at above 100~GHz band, where a carrier signal with a high signal-to-noise ratio is desired to achieve higher data rates.
In this study, we demonstrate two simple and efficient architectures of wireless links based on a microresonator comb.
We show experimentally that simultaneous modulation and detection of multiple comb lines result in $>10$ times stronger modulation signal strength than two-line detection at a receiver.
Successful transmission of complex modulation format up to 64 quadrature amplitude modulation proves that a microresonator comb and the proposed modulation method are effective in modern wireless communication.
\end{abstract}

\section{Introduction}
In response to the continued exponential growth of the data traffic in wireless networks \cite{5gwireless}, increasing efforts have been devoted to wireless communication technologies above the 100~GHz band, where abundant frequency resources exist \cite{nagatsuma2016advances}.
Four frequency bands with a total bandwidth of 137~GHz between 275~GHz and 450~GHz were identified for implementing land mobile and fixed service applications a few years ago \cite{WRC19_final}.
Data rates of 100~Gbit/s have been achieved at the frequency ranges by employing quadrature amplitude modulation (QAM) such as 16-QAM  \cite{koenig2013wireless,nagatsuma2016real,chinni2018single, hamada2018300}.
A simple approach to further progress in data rates while keeping the same bandwidth occupancy is exploiting higher modulation orders.
So, higher power and lower white noise floor in carrier signals are demanded since a larger signal-to-noise ratio (SNR) in detected signals is required for that \cite{Chen2018Influence}. 

There are roughly two generation and modulation/demodulation methods of high-frequency radio frequency (RF) waves with low phase noise: all-electronics and photonics-based approaches. 
The former completes all processes of signal handling in electronic circuits. 
It is one step ahead of the photonics-based counterpart regarding monolithic integrability, which is a key factor for mass production and reducing production costs. 
However, multi-stage frequency multiplication and amplification processes required for producing high-frequency RF waves intrinsically degrade their phase noise and SNR, posing a limitation for the maximum data rate transmitted.
Modulation is usually carried out by mixing carrier and modulated signals.
On the other hand, a photonics-based approach is suitable for generating a lower-phase-noise wave.
It produces a high-frequency RF wave via heterodyne detection of optical lines with a fast photodiode (PD) \cite{nagatsuma2013terahertz}.
The phase noise is not limited by a frequency of a generated wave but by the relative phase noise of seed light, which can be controlled independently of the photo-mixing process. 
In addition, it is advantageous that signal encoding is possible in the optical domain by using sophisticated modulation schemes in optical communication.
Pure electronic or optoelectronic devices carry out the signal reception \cite{harter2019wireless}.
On the other hand, systems to achieve ultra-low phase noise tend to be bulky \cite{li2019low}.
Also, the output power is limited by the maximum currents a PD can handle and the PD's responsivity.

A Kerr microresonator frequency comb~\cite{herr2014temporal,brasch2016photonic}, or microcomb, is a potential candidate to address the challenges in conventional photonics-based oscillators. 
Its repetition frequency is typically located between 10~GHz and 1~THz, and the generation of RF waves with frequencies higher than 100~GHz has already been demonstrated via direct detection of the comb lines with ultra-fast PDs \cite{zhang2019terahertz, huang2017globally, tetsumoto2020300, wang2021towards, tetsumoto2021optically}.
There are three main advantages of using the microcomb in wireless communication. 
(i) It can enhance detected modulation signal power thanks to constructive interference of multiple RF waves generated from multiple comb lines, as observed for carrier power in some studies~\cite{kuo2010spectral,wang2021towards}. 
It will help to gain higher SNR out of the limited photocurrent of a PD. 
(ii) The system can be more compact and simple than other photonics-based oscillators. 
Aside from being on-chip, it will eliminate the need for some optical components required for other photonic systems (e.g., optical couplers and spectrum filters). 
The large mode spacing allows a microcomb to be modulated and detected directly, as demonstrated in this study. 
(iii) Its phase noise can be reduced significantly through stabilization to micro-wave references or optical fibers~\cite{zhang2019terahertz,tetsumoto2020300,kuse2022low}, optical frequency division \cite{tetsumoto2021optically}, dispersion engineering of microresonators \cite{stone2020harnessing}, or turning experimental parameters~\cite{yi2017single,tetsumoto2021effects}. 
In fact, it has enabled demonstrating the record-low phase noise at 300~GHz \cite{tetsumoto2021optically}. 

In this study, we demonstrate the unique features of a microcomb as a transmitter/receiver in wireless links.
The three advantages mentioned above are addressed, respectively, as follows.
The first one is demonstrated experimentally by employing power-equalized comb lines.
We observe more than 10~dB gain of modulation signal strength when multiple optical lines are detected compared to a two-line case.
We perform wireless communication experiments with a microcomb-based 300~GHz transmitter and receiver to show the second point. 
As a result, successful transmission of up to 64-QAM signal is confirmed, whose limitation is not given by the microcomb.
The transmitter and receiver are stabilized to a common reference during the experiments, which corresponds to the condition that the third advantage is in effect.
This study will be a milestone in developing compact microcomb-based photonic millimeter- and terahertz-wave transceivers and help achieve higher data rates in future wireless communication systems.

\section{Operation principle}
We modulate the whole comb spectrum and detect it directly in this study.
By doing so, the detected modulation signal power is enhanced compared to two-line heterodyne detection under the same photocurrent. 
This effect has been observed in previous studies for carrier signal power~\cite{kuo2010spectral,wang2021towards}.

\begin{figure*}[!ht]
    \centering
    \includegraphics[width=\linewidth]{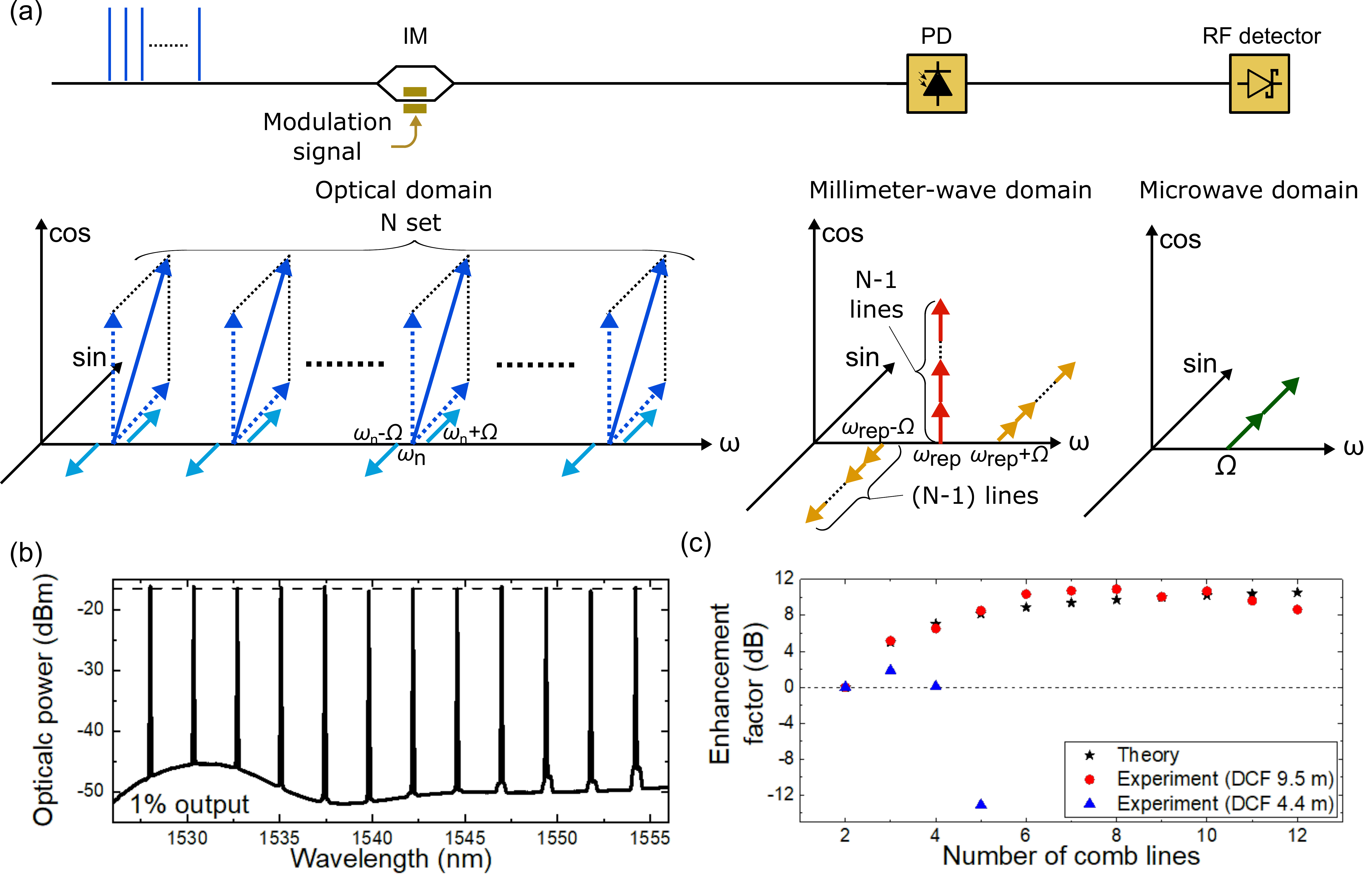}
    \caption{\footnotesize{ (\textbf{a}) Schematic illustration of operation principle. The upper half shows an experimental situation considered, and the lower half is the spectrum of signals at each stage. (\textbf{b}) Equalized comb lines from a 1~\% monitor port after the amplification. The power difference between combs is less than $\pm 0.5$~dB (dashed line). (\textbf{c}) Measured enhancement of detected modulation signal strength. The dashed line shows 0~dB of enhancement.}}
    \label{fig1}
\end{figure*}

Power spectrum of photocurrent from a PD of $|I_\mathrm{PD}(\omega)|^2$ is given by,
\begin{equation}
    |I_\mathrm{PD}(\omega)|^2 = |G(\omega ) \cdot P(\omega )|^2,
    \label{eq1}
\end{equation}
where $\omega$ is angular frequency, $|G(\omega )|^2$ and $|P(\omega )|^2$ are power spectra of impulse response of a PD and an input wave, respectively.
So, the products of the PD output will be proportional to the square of the input wave power spectrum, assuming the PD's response is homogeneous and linear over the frequency of interest.
Figure~\ref{fig1}(a) illustrates the modulation/detection methods and spectra obtained at the respective stages.
The input is power-equalized $N$ comb lines with amplitude of $1/\sqrt{N}$ normalized so that total power becomes 1. 
They are modulated with an intensity modulator (IM) with a modulation signal $\phi _\mathrm{RF} = M \sin{\Omega t }$, where $\Omega$ is an RF frequency, and $M$ is modulation depth. 
There is no in-balance in the optical splitting and combining processes in the IM, and the light only on one arm experiences the modulation. 
Then, the output of the modulator $E_\mathrm{out}$ is \cite{urick2015fundamentals},
\begin{eqnarray}
    & E_\mathrm{out} &= \sum_{n=1}^{N} \frac{1}{\sqrt{N}}\left[ \cos{\omega _{n} t} 
    + \sum_{k=-\infty}^{\infty} J_{k}(M)\sin\{(\omega _{n} + k\Omega )t \} \right] \nonumber \\ 
    &\sim & \sum_{n=1}^{N} \frac{1}{\sqrt{N}}[ \cos{\omega _{n} t}
    - J_{1}(M)\sin\{(\omega _{n} - \Omega )t \} + J_{0}(M)\sin{\omega _{n}t} + J_{1}(M)\sin\{(\omega _{n} + \Omega )t \} ] \nonumber \\ 
    &=& \sum_{n=1}^{N} \frac{1}{\sqrt{N}} A_\mathrm{n},
    \label{eq_opt}
\end{eqnarray}
where $n$ and $k$ are integers, $\omega _{n}$ is the angular frequency of $n$~th comb mode, $J_k$ is $k$~th-order Bessel function of the first kind, and $A_\mathrm{n}$ is the sum of the center and sideband modes around $n$-th optical line.
The factor $1/2$, which explains power splitting and combining, is omitted, and the phase difference between the two arms of $\pi/2$ is assumed for simplicity.
Higher order sidebands are neglected since the modulation depth $M$ is small, and $J_{-k}=(-1)^{k}J_{k}$ is employed.
The left bottom of Fig.~\ref{fig1}(a) depicts the optical spectrum of the comb lines at the IM output. 
These optical lines are sent to a PD and generate a carrier and its sidebands (modulation tones) in the millimeter-wave domain through $I_\mathrm{PD} \propto |E_\mathrm{out}|^2$.
The millimeter-wave carrier with the angular repetition frequency $\omega _\mathrm{rep}=\omega _{n} - \omega _{n-1}$ and its sidebands are generated through the interference between adjacent optical lines and their sidebands of,

\begin{equation}
    I_\mathrm{adj} = \frac{1}{N} \sum_{n=1}^{N-1}  A_{n}A_{n+1}.
    \label{eq_components1}
\end{equation}

\noindent
Each term of Eq.~\ref{eq_components1} gives the following components for the respective angular frequencies when the optical lines are phase-coherent:

\begin{eqnarray}
    \omega _\mathrm{rep} &:& \frac{1}{2N} \{ 1 - J_{0}^{2}(M) - 2J _{1}^{2}(M) \} \cos {\omega _\mathrm{rep}t}, \nonumber \\
    \omega _\mathrm{rep} - \Omega &:& -\frac{J_{1}(M)}{N} \sin\{ \left(\omega _\mathrm{rep} - \Omega \right)t\}, \nonumber \\
    \omega _\mathrm{rep} + \Omega &:& \frac{J_{1}(M)}{N}  \sin\{ \left(\omega _\mathrm{rep} + \Omega \right)t\}. \nonumber
    \label{eq_components2}
\end{eqnarray}

\noindent
The resulting spectrum around the carrier is schematically shown in the center bottom of Fig.~\ref{fig1}(a), where higher-order sidebands are neglected because they are small in the weak modulation condition considered.
The amplitude of each frequency component is multiplied by $N-1$ owing to the contribution from each combination in Eq.~\ref{eq_components1}.
The envelope of the generated millimeter-wave can be captured with an RF detector, such as a Schottky barrier diode (SBD). 
The modulation signal $P_\mathrm{mod} (N)$ with the angular frequency of $\Omega$ is detected in the microwave domain via the interaction between the carrier and the two sidebands, which is expressed as,

\begin{eqnarray}
P_\mathrm{mod} (N) &\propto & 2 \times \frac{N-1}{2N} \{ 1 - 2J _{1}(M)^{2} - J_{0}(M)^{2} \} \times  J_{1}(M)\frac{N-1}{N} \times \frac{\sin{\Omega t}}{2} \nonumber \\ 
&=&  \frac{1}{2} \left( \frac{N-1}{N} \right) ^{2} \{ 1 - 2J _{1}^{2}(M) - J_{0}^{2}(M) \} J_{1}(M) \sin{\Omega t}.
\label{eq_mod}
\end{eqnarray}

\noindent
Therefore, the enhancement factor of the modulation signal's power spectrum owing to the multiple comb lines can be expressed as follows:
\begin{equation}
    \left| P_\mathrm{mod}(N) \right| ^2 \propto \left| \left( \frac{N-1}{N} \right) ^{2} \right| ^{2} = \left( \frac{N-1}{N} \right) ^{4} .
    \label{eq_enhance}
\end{equation}
So, the modulation signal power will be enhanced by $|P_\mathrm{mod}(N)|^2/|P_\mathrm{mod}(2)|^2=16\left( \frac{N-1}{N}\right)^{4}$ compared to two-line cases, which will become 16 times (12~dB) when $N \to \infty$.
This number is from the 6~dB respective power enhancement of the carrier and sidebands in the millimeter-wave domain.
The key point is that each adjacent comb line pair produces an identical set of RF waves regarding amplitudes, frequencies, and phases at the detection.
Thus, the principle will be valid for modulation methods that gives the carrier and its sideband signals via photo-detection as long as the optical comb lines are phase-locked and the generated waves interfere constructively.

We demonstrate a simple experiment to confirm the effect. 
We prepare microcomb lines with a 300~GHz frequency spacing and equalized power by using a waveshaper and an erbium-doped fiber amplifier (EDFA), as shown in Fig.~\ref{fig1}(b).
They are modulated by a 1.5~GHz sinusoidal wave with an IM and sent to an unitravelling-carrier photodiode (UTC-PD).
The envelope of the generated 300~GHz wave is detected with an SBD.
A dispersion compensating fiber (DCF, DCF-38 Thorlabs with $\sim $-38~ps/nm$\cdot$km) is inserted between the waveshaper and the UTC-PD to compensate for the dispersion effect.
We change the number of comb lines to inject, and record the change in the detected signal power of the modulation tone.
The photocurrent of the UTC-PD is kept at 0.5~mA for all measurements.
Figure~\ref{fig1}(c) shows the results measured with two different DCF lengths, where they are normalized by the power at $N=2$.
With the DCF of 9.5~m (red circle plots), the power of the detected modulation signal increases as the number of comb lines increases, and the trend is close to the theoretical expectation (black star plots).
The enhancement factor of as high as 10.9~dB is obtained with 7 comb lines.
On the other hand, the signal strength does not go up so high with increased numbers of comb lines but drops significantly with a 4.4~m DCF (blue triangular plots, at $N=5$), where the dispersion of the optical path is not compensated correctly.
This presents that the multiple comb line detection scheme can cause a reverse effect if the phase of each comb line is not appropriately aligned~\cite{wang2021towards}.
The small effect of the dispersion also can be seen in the plots for 9.5~m DCF at $N>9$.

\section{Microcomb generation, stabilization \& characterization}
\begin{figure*}[!ht]
    \centering
    \includegraphics[width=\linewidth]{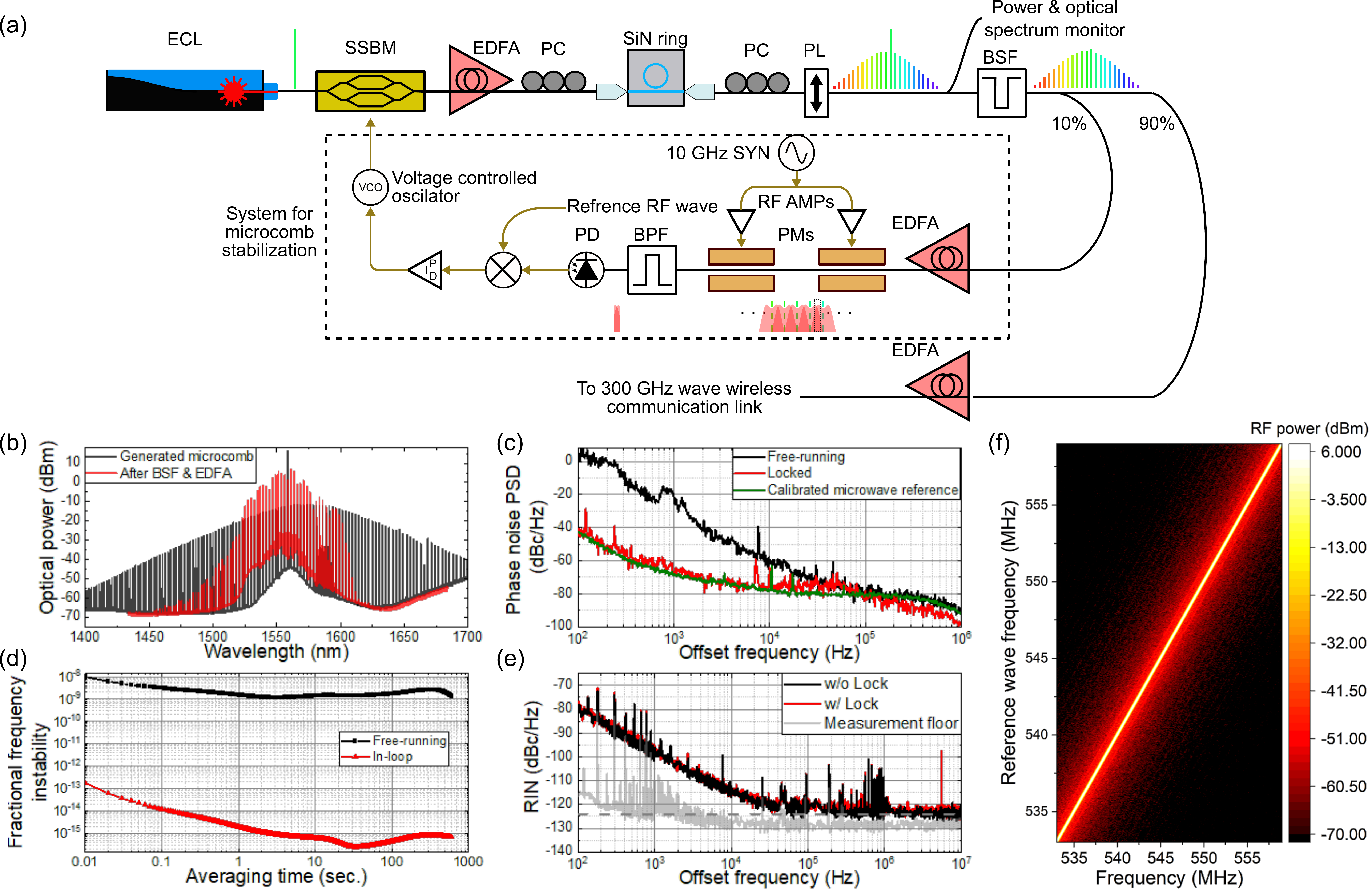}
    \caption{\footnotesize{ (\textbf{a}) Schematic drawing of the experimental setup for microcomb generation and stabilization. See text for details. Generated microcomb's (\textbf{b}) optical spectra, (\textbf{c}) phase noise, (\textbf{d}) frequency instability, and (\textbf{e}) relative intensity noise. The measurement floor in (e) (shaded curve) is obtained when no light is sent into the UTC-PD. (\textbf{f}) Result of the repetition frequency tuning.}}
    \label{fig2}
\end{figure*}

Figure~\ref{fig2} depicts an experimental setup for microcomb generation and stabilization. Single sideband modulation is applied to continuous-wave (CW) light from an external cavity laser diode (ECL, Toptica CTL1550) with a single sideband modulator (SSBM), and the output is amplified with an EDFA. 
The light is coupled with a ring resonator made of silicon nitride (SiN) after polarization alignment using a polarization controller (PC). 
A soliton microcomb is excited by fast scanning the strong pump light with the SSBM \cite{briles2018interlocking, kuse2019control}. 
The generated microcomb is polarized with another PC and a polarizer (PL). We employ polarization-maintaining fibers after here. 
A small portion of the comb light is sampled for power and spectrum monitoring during the experiment. 
The pump light is suppressed with a bandstop filter (BSF), and the output is split into two paths with a 90:10 optical coupler. 
The 90~\% of the light is amplified with an EDFA and sent to a wireless link, explained later. 
Figure~\ref{fig2}(b) shows the optical spectra of the generated microcomb and the filtered and amplified comb on the 90~\% port.
The bandwidth of the filtered comb spectrum is mainly limited by the BSF and EDFA gain bandwidths. 
Pump-to-comb conversion efficiency is estimated to be $\sim 4$~\% from the ratio between the pump power and the total power of the other combs in the optical spectrum.

The output from the 10~\% port is used to stabilize the microcomb through the method described in \cite{tetsumoto2020300}. 
Although the suppression of the phase noise and frequency drift is not required for the communication experiments in this study based on self-homodyne detection, we implement this to show the microcomb's capability of being synchronized to a reference signal.
For the stabilization, sidebands of the comb lines are generated with two cascaded phase modulators (PMs) driven by a 10~GHz wave from a synthesizer (SYN), whose RF power is amplified to about 1~W with RF amplifiers (RF AMPs). 
At around a center frequency of two comb lines, the 15~th sidebands from each comb are closely located thanks to the broad bandwidth of the generated electro-optic (EO) combs. 
The spectrally overlapped frequency region is sampled with a bandpass filter (BPF), and the output is detected with a photodiode (PD). 
The detected beat frequency $f_\mathrm{beat}$ is described as,
\begin{equation}
     f_\mathrm{beat} = (f_\mathrm{n+1} - 15 f_\mathrm{10G}) - (f_\mathrm{n} + 15 f_\mathrm{10G}) = f_\mathrm{rep} -  30 f_\mathrm{10G}.
    \label{eq3}
\end{equation}
where $f_\mathrm{n+1}$ and $f_\mathrm{n}$ are frequencies of $n+1$~th and $n$~th microcomb modes, and $f_\mathrm{10G}$ is frequency of the 10~GHz SYN.
So, the phase noise of the beat signal is,
\begin{equation}
     \delta _{f_\mathrm{beat}} = \delta _{f_\mathrm{rep}} -  30 \delta _{f_\mathrm{10G}}.
    \label{eq4}
\end{equation}
where we employ a notation of $\delta _{f}$ for the phase noise of a signal with frequency $f$.
We produce an error signal by mixing the beat signal with an RF reference and generate a control signal via a proportional integral derivative (PID) loop filter. 
The control signal drives a voltage-controlled oscillator, whose output is applied to the frequency control channel of the SSBM. 
Note that both the 10~GHz SYN and the reference RF wave in the system are synchronized to the same clock signal; thus, the microcomb is synchronized as well. 
Figure~\ref{fig2}(c) presents the measured phase noise. 
The free-running comb noise is suppressed to the calibrated noise level of the 10~GHz reference SYN in the locked condition. 
The feedback bandwidth is set to slightly lower than the cross-point frequency of the free-running and reference noise of 100~kHz so that the noise after the stabilization follows the lower of them.
The phase noise at 10~kHz offset is -64~dBc/Hz (ignoring the spurious peak from the SYN, the noise level is $-75$~dBc/Hz). 
The details of the phase noise measurement are shown in Supplementary Material. 
Figure~\ref{fig2}(d) presents the frequency instability of the microcomb and the in-loop signal evaluated in terms of modified Allan deviation. 
The obtained frequency instability at 1~second is $1.5\times 10^{-9}$ for the free-running microcomb and $1.9\times 10^{-15}$ for the in-loop signal, respectively. 
The latter confirms that the microcomb is stabilized to the reference tightly.
In addition, we measure the relative intensity noise (RIN) of the generated 300~GHz wave as shown in Fig.~\ref{fig2}(e). 
The number of comb lines and optical path dispersion are controlled in the same way as Fig.~\ref{fig1}(c), where $N=10$, the DCF of 9.5~m and the UTC-PD photocurrent of 6~mA are employed. 
The RIN reaches a low noise floor of -124~dBc/Hz (grey dashed line) in both locked and unlocked conditions (black and red curves).

Also, we can use the stabilization scheme for fine-tuning the repetition frequency of the microcomb.
Equation~\ref{eq3} can be rewritten as,
\begin{equation}
    f_\mathrm{rep} = 30 f_\mathrm{10G} + f_\mathrm{beat}.
    \label{eq5}
\end{equation}
This indicates that $f_\mathrm{rep}$ can be tuned by controlling the reference RF wave frequency to which $f_\mathrm{beat}$ is stabilized. Note that changing $f_\mathrm{10G}$ will only shift $f_\mathrm{beat}$ not $f_\mathrm{rep}$. 
To demonstrate this, we acquire the RF spectrum of the beat signal and observe its transition when we change the reference RF frequency in steps. Figure.~\ref{fig2}(f) shows the result. 
The beat frequency follows the reference frequency linearly, and the tuning range of 26~MHz is obtained without losing the lock. 
Though the frequency sweep is performed slowly (i.e., about 100~kHz/sec), which is limited by the data acquisition time for the measurement, the scanning speed can be faster as long as the phase lock loops can catch up.

\section{Wireless communication experiments}
\begin{figure*}[!ht]
    \centering
    \includegraphics[width=\linewidth]{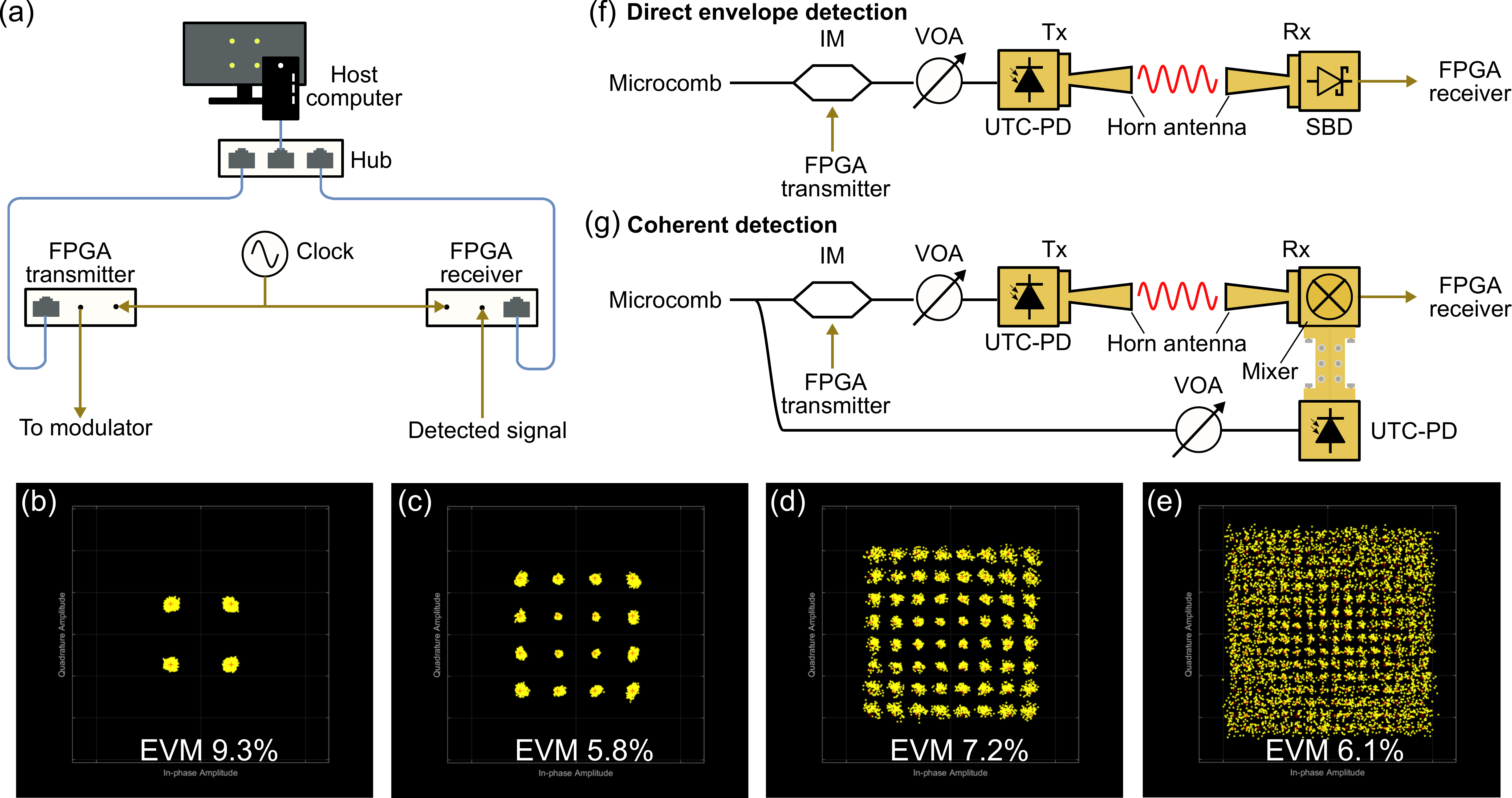}
    \caption{\footnotesize{(\textbf{a}) Schematic drawing of FPGA-based modulation/demodulation system. \textbf{b})-(\textbf{e}) Constellation diagrams for 4-QAM, 16-QAM, 64-QAM and 256-QAM modulation schemes, respectively, in FPGA direct link. Schematic drawing of wireless links based on (\textbf{f}) direct envelope detection and (\textbf{g}) coherent detection. See text for details. }}
    \label{fig3}
\end{figure*} 

Wireless communication systems in this study consist of a software part to produce and analyze digital signals and a hardware part to modulate, transmit and receive 300~GHz waves.
Digital signal processing (DSP) in the wireless system is performed with MATLAB \& Simulink (details are in Supplementary Material), and transmitter and receiver based on field programmable gate arrays (FPGAs) (USRP N210, Ettu) acting as interfaces between the software and the hardware parts. 
Figure~\ref{fig3}(a) depicts the schematic of the system.
The host computer, which processes the Simulink program, the FPGA transmitter, and the receiver, are connected via Ethernet cables and hubs.
The transmitter and the receiver share the same clock signal as the microcomb.
The software program generates modulation signals encoded with either 4-QAM, 16-QAM, 64-QAM, or 256-QAM. 
The center frequency and the symbol rate of the signal are set to 1.5~GHz and 200~kHz, respectively, where the rather low symbol rate is to avoid overrun and underrun in the software-hardware communication.
The transmitter is configured to generate the produced modulation signal as an arbitrary waveform generator.
The modulated signal is applied to an intensity modulator (IM) in the hardware system.
The in-phase and quadrature-phase components of the signal are superimposed in the RF wave domain.
The detected signal in the hardware is down-converted to baseband frequency, digitized in the FPGA receiver, and analyzed with the Simulink program in the host computer in real time. We evaluated the received signal's root mean square (RMS) error vector magnitude (EVM).

Figure~\ref{fig3}(b)-(e) shows the results of a back-to-back preliminary experiment, where the FPGA transmitter and the FPGA receiver are connected with a 1.5~m electric cable and a 10~dB attenuator. 
Constellation points can be recognized clearly in 4-QAM, 16-QAM, and 64-QAM cases with EVM of 9.3~\%, 5.8~\%, and 7.2~\%, respectively, whereas 256-QAM constellation with EVM of 6.1~\% is not resolved well. These results pose the lowest levels of the reachable EVM in the following experiments.

We tested two types of wireless links. 
One is based on direct envelope detection of the 300~GHz carrier with an SBD (Fig.~\ref{fig3}(f)). 
The stabilized microcomb is modulated by the IM and detected with a UTC-PD, where a 300~GHz carrier is generated. 
The 300~GHz wave is radiated into free space with a horn antenna, travels 0.1~m, is captured with another horn antenna, and is detected with the SBD. 
The output from the detector is sent to the FPGA receiver. 
The other system employs coherent mixing of two 300~GHz waves (Fig.~\ref{fig3}(g)). 
The modulated wave is generated in the same way as the direct envelope detection.
Another wave for the down-conversion is produced by detecting the same microcomb before the modulation in this proof-of-concept study.

The generated two waves are mixed with a 300~GHz fundamental mixer, and the intermediate frequency (IF) signal is supplied to the FPGA receiver.
Although a homodyne detection system requires precise control of the relative phase of two waves detected in general, its fluctuation was slow enough in our experiment owing to the short path length after the fork, and no active control was implemented.
Note that we omit a DCF in the two setups since we can obtain sufficient SNR for the demonstration without it.

\begin{figure*}[!ht]
    \centering
    \includegraphics[width=\linewidth]{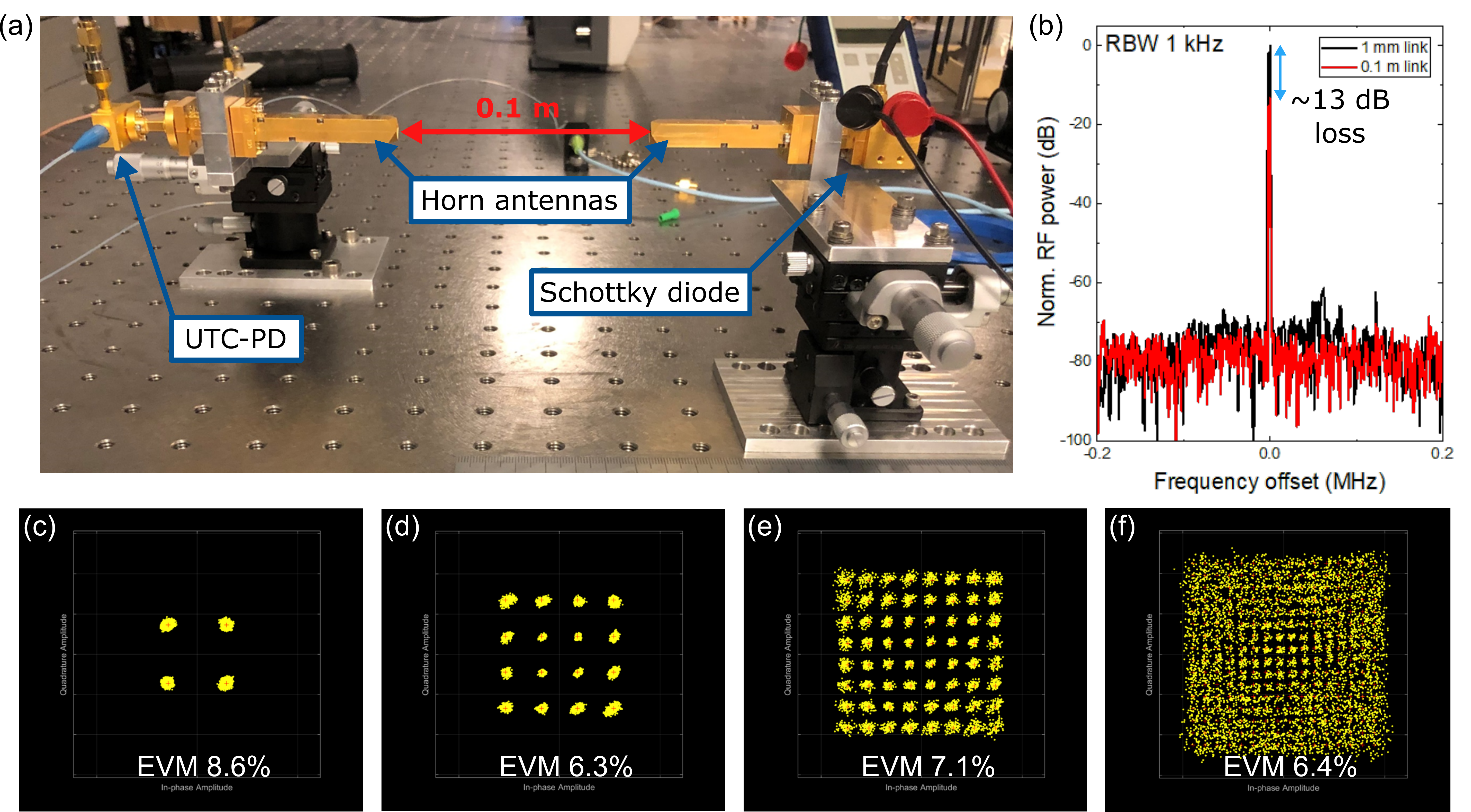}
    \caption{\footnotesize{(\textbf{a}) A picture of the wireless link based on direct envelope detection. (\textbf{b}) RF spectra of detected 1.5~GHz modulated tones in transmission loss test. RBW is 1~kHz. (\textbf{c})-(\textbf{f}) Constellation diagrams for 4-QAM, 16-QAM, 64-QAM and 256-QAM modulation schemes, respectively, in direct envelope detection.}}
    \label{fig4}
\end{figure*}

\begin{figure*}[!ht]
    \centering
    \includegraphics[width=\linewidth]{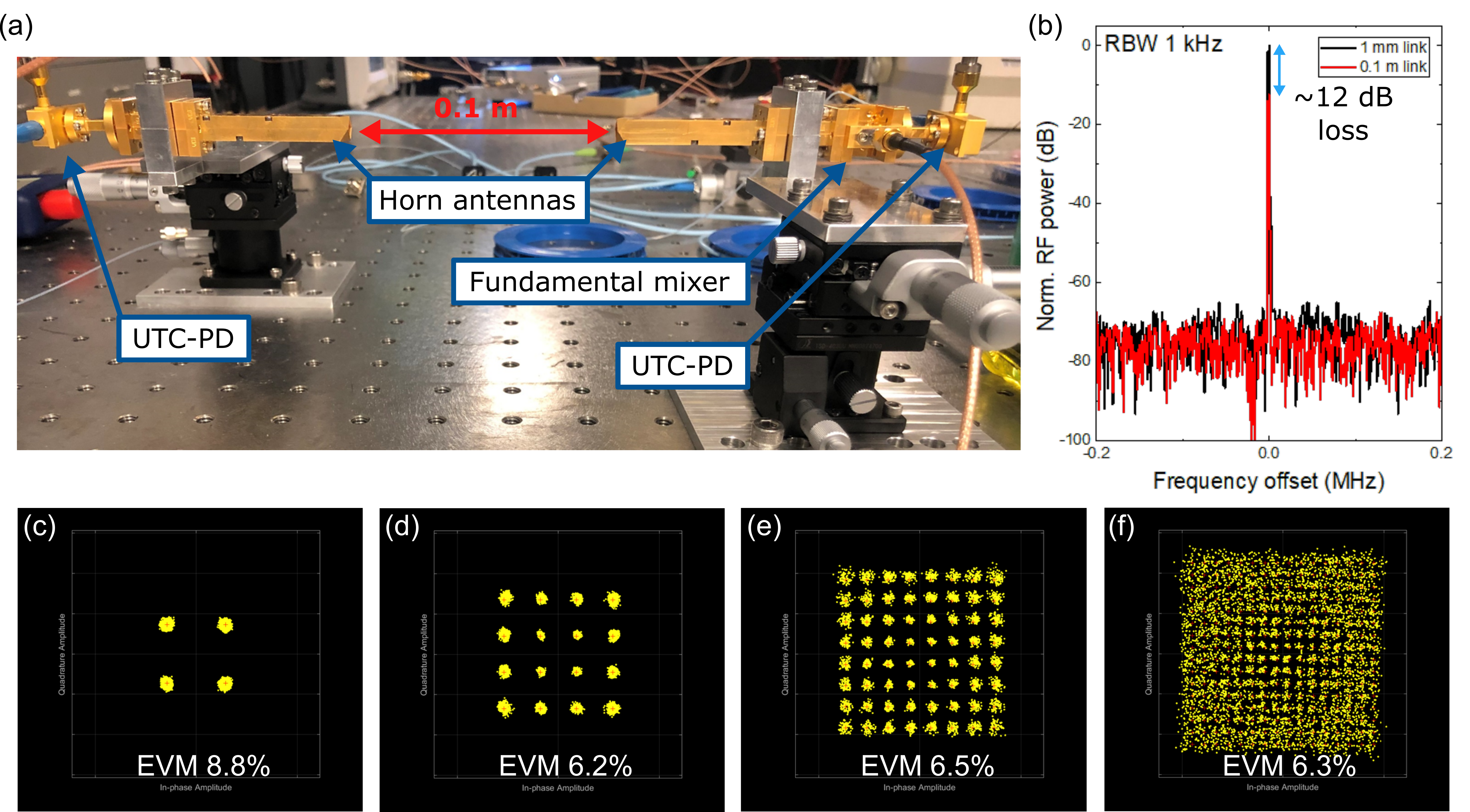}
    \caption{\footnotesize{(\textbf{a}) A picture of the wireless link based on coherent detection. (\textbf{b}) RF spectra of detected 1.5~GHz modulated tones in transmission loss test. RBW is 1~kHz. (\textbf{c})-(\textbf{f}) Constellation diagrams for 4-QAM, 16-QAM, 64-QAM, and 256-QAM modulation schemes, respectively, in coherent detection.}}
    \label{fig5}
\end{figure*}

Figure~\ref{fig4}(a) presents a picture of the actual wireless link based on direct envelope detection. We employ horn antennas with a designed gain of 26~dBi. The transmittance is maximized by controlling the positions of the antennas with 3-axis mechanical stages. To check the transmission loss, we detect the 1.5~GHz sinusoidal wave modulation signal delivered by a 300~GHz carrier in the 0.1~m link and a 1~mm link. Figure~\ref{fig4}(b) shows the RF spectra of the detected signals. We observe about 13~dB reduction of the SNR through the 0.1~m free-space trip compared to the 1~mm link.
Figure~\ref{fig4}(c)-(f) displays the results of the wireless communication. The obtained EVM are 8.6~\%, 6.3~\%, 7.1~\%, and 6.4~\% for 4-QAM, 16-QAM, 64-QAM, and 256-QAM, respectively. So, no apparent degradation of signal quality is observed compared to the direct FPGA link experiment (Fig.~\ref{fig3}(b)-(e)). 
The wireless link for the coherent detection is presented in Fig.~\ref{fig5}(a). Most components are the same as ones in the direct envelope detection system, but the SBD is replaced with a fundamental mixer, and another UTC-PD is attached to it to supply a 300~GHz wave to the LO port. Again, we demonstrate the transmission test and observe a 12~dB decrease in the SNR in the 0.1~m link. The acquired constellation diagrams are shown in Fig~\ref{fig5}(c)-(f). EVM of 8.8~\%, 6.2~\%, 6.5~\%, and 6.3~\% are obtained for 4QAM, 16QAM, 64QAM, and 256QAM, respectively, which are merely limited by the FPGA transmitter and receiver.

\begin{figure*}[!ht]
    \centering
    \includegraphics[width=\linewidth]{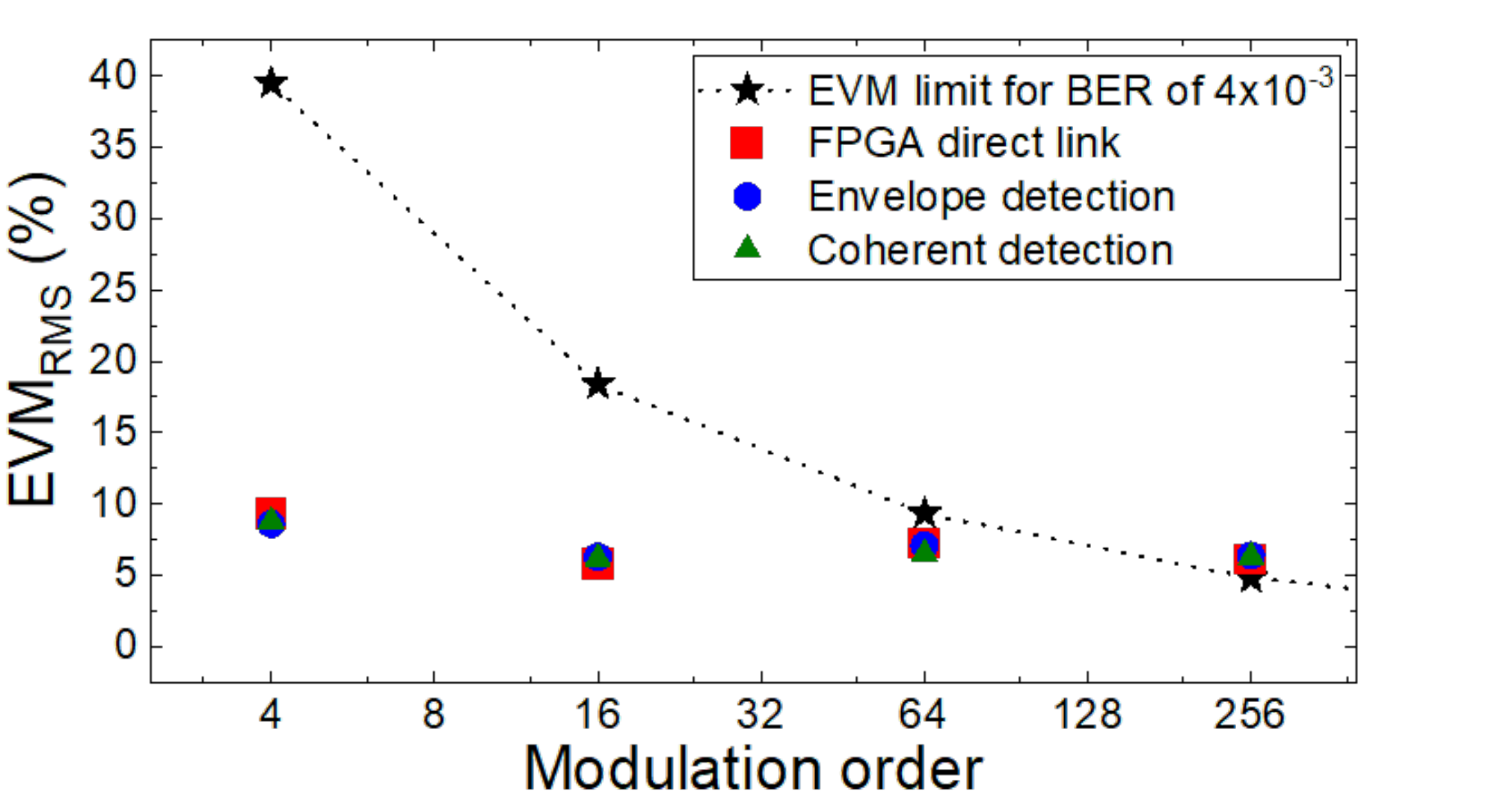}
    \caption{Summary of obtained EVM in wireless communication experiments}
    \label{fig6}
\end{figure*}

Figure~\ref{fig6} summarises the results of the communication experiments. 
EVM threshold to obtain bit error ratio (BER) of $4\times 10^{-3}$ is displayed by the black dashed line with star symbols as a measure of successful transmission \cite{shafik2006extended,chinni2018single}. With the EVM level, the BER will be decreased to $10^{-15}$ level by the forward error correction with 7~\% overhead \cite{chang2010forward}.
The obtained EVM levels are below the EVM limit in the experiment up to 64-QAM but slightly beyond the limit in the 256-QAM demonstration.
The limitation is given by the FPGA transmitter and receiver (phase noise is -80~dBc/Hz at 10~kHz offset for a 1.8~GHz signal according to their specification sheet).
This is convincing because the microcomb's phase noise is canceled out and we will see only modulation signal noise at the reception in the two wireless links demonstrated in this study.

\section{Discussion \& Outlook}
Here, we discuss microcomb's advantages for wireless communication and future perspective in more detail.

Our experiments demonstrate that a microcomb can be an oscillator in wireless links utilizing complex modulation formats.
The system is relatively simple compared to other photonics-based approaches because it starts from a single CW light, and no spectral filtering process is required, which allows comb lines to propagate in a shared path all the time.
This is, in particular, significant difference from photonics-based systems using conventional comb sources, which suffer from large loss due to splitting, filtering and combining processes and sometimes even require mechanisms to compensate fluctuation of path length difference \cite{nagatsuma2013terahertz}.

Also, the employed two architectures will benefit from respective merits as follows.
The direct envelope detection scheme does not require low-phase-noise carriers because the carrier noise will be canceled out at the detection, whereas its sensitivity is limited by a detector employed.
So, the critical parameters for this configuration are the noise property of RF sources in modulation/demodulation (e.g., arbitrary waveform generator for encoding, an oscillator for decoding), signal strength at a transmitter, and detector sensitivity.
A microcomb will contribute to improving the signal strength via the simultaneous modulation and detection of comb lines.
The coherent detection scheme will enhance detection sensitivity by a strong LO signal and give a larger bandwidth at the IF output in general.
However, detected SNR will be affected by the phase noise of both RF and LO signals injected into a mixer.
Also, the RF and LO signals need to have a small frequency drift or, ideally, share the common clock signals to help carrier frequency and phase synchronization in the DSP work properly.
Regarding the phase noise, a 300~GHz microcomb can be comparable (-100~dBc/Hz at 10~kHz) to an arbitrary waveform generator at 10~GHz used in a cutting-edge communication experiment \cite{hamada2018300} (-95~dBc/Hz at 10~kHz offset, Keysight M8196A \cite{keysight_M8196A}) through the optical frequency division technique~\cite{tetsumoto2021optically}.
So far, white phase noise floors of microcombs at 300~GHz are observed at around -120~dBc/Hz limited by shot noise and noise figure of electronic devices \cite{zhang2019terahertz,tetsumoto2021optically}. Note that this is still $>10$~dB lower than that of a typical electronic 300~GHz oscillator based on frequency multiplication \cite{dan2020superheterodyne}.
In addition, a microcomb oscillator can address the clocking problem and provide robust operation, as demonstrated in this study.

Two crucial challenges need to be addressed to realize low-cost and portable microcomb-based transceivers for future wireless communication: integrating the system and reducing power requirement.
Although a system based on the optical fiber connection like the one in this study can be packaged compactly, making it on-chip is more desirable for its miniaturization and mass production.
Such a platform will also mitigate dispersion effects, observed in Fig.~\ref{fig1}(e), thanks to much shorter path lengths.
Hybrid and heterogeneous integration \cite{liang2021recent,kaur2021hybrid} will be a key technology to achieve the goal since active components like lasers and media for nonlinear optical effects are usually made of different materials with different bandgap structures.
Recently, soliton generation has been demonstrated with InP/Si semiconductor lasers monolithically integrated with high $Q$ SiN resonators \cite{xiang2021laser}.
Regarding subsystems, the broadband EO comb we employed to stabilize the comb's repetition frequency may not be a suitable option when pursuing system integration as it requires high-power RF amplifiers.
Phase-locking the 300~GHz comb to a lower-frequency microcomb at the microwave domain can be an alternative way to down-convert and stabilize the comb frequency \cite{spencer2018optical,liu2020photonic}.
On-chip low half-wave voltage modulators may become a straightforward solution to relax the RF power requirement \cite{zhang2021integrated}.

The excitation power requirement for the microcomb generation has been lowered through progress in fabrication technologies and theoretical and experimental understanding of its generation physics.
For example, a soliton microcomb has been excited with 100-mW-level electrical power (1-mW-level optical power) by employing an ultrahigh $Q$ SiN ring resonator ($Q \sim 10^{7}$)\cite{stern2018battery}.
Pump-to-comb conversion efficiency is also an important measure to evaluate the energy cost.
It can be controlled via various parameters such as the resonator's size and loss and coupling between a waveguide and a resonator, where the efficiency close to 10~\% is predicted for a 300~GHz SiN resonator with intrinsic $Q$ in the order of $10^6$ under a highly over-coupled condition~\cite{jang2021conversion}.
To achieve the efficiency of $>10$~\%, using dark pulses or coupled resonator configurations can be effective approaches~\cite{xue2017microresonator,kim2019turn, xue2019super,helgason2022power}.

In summary, we demonstrated 300~GHz wireless communication links based on a microcomb for the first time.
The simultaneous modulation of the multiple microcomb lines enhanced the detected modulation signal strength by more than 10~dB compared to the two-line case.
We performed a transmission experiment of complex modulation format in two types of wireless links by using a microcomb stabilized to a reference clock.
No degradation of the signal quality is observed in the experiments, and the QAM signal as high order as 64 was transmitted successfully in both links.
The results show that a microcomb is a promising candidate as a photonic oscillator that boosts data capacity in future wireless communication.

\bibliography{MicrocombWireless}

\begin{thebibliography}{10}
\newcommand{\enquote}[1]{``#1''}

\bibitem{5gwireless}
\enquote{{5G Wireless: Capabilities and Challenges for an Evolving Network},}
  \url{https://www.gao.gov/assets/gao-21-26sp.pdf}.

\bibitem{nagatsuma2016advances}
T.~Nagatsuma, G.~Ducournau, and C.~C. Renaud, \enquote{Advances in terahertz
  communications accelerated by photonics,} {\protect\JournalTitle{Nature
  Photonics}} \textbf{10}, 371--379 (2016).

\bibitem{WRC19_final}
\enquote{{World Radiocommunication Conference 2019 (WRC-19) Final Acts},}
  \url{https://www.itu.int/dms_pub/itu-r/opb/act/R-ACT-WRC.14-2019-PDF-E.pdf}.

\bibitem{koenig2013wireless}
S.~Koenig, D.~Lopez-Diaz, J.~Antes, F.~Boes, R.~Henneberger, A.~Leuther,
  A.~Tessmann, R.~Schmogrow, D.~Hillerkuss, R.~Palmer \emph{et~al.},
  \enquote{Wireless sub-{THz} communication system with high data rate,}
  {\protect\JournalTitle{Nature photonics}} \textbf{7}, 977--981 (2013).

\bibitem{nagatsuma2016real}
T.~Nagatsuma, Y.~Fujita, Y.~Yasuda, Y.~Kanai, S.~Hisatake, M.~Fujiwara, and
  J.~Kani, \enquote{Real-time 100-{Gbit/s} {QPSK} transmission using
  photonics-based 300-{GHz}-band wireless link,} in \emph{2016 IEEE
  International Topical Meeting on Microwave Photonics (MWP),}  (IEEE, 2016),
  pp. 27--30.

\bibitem{chinni2018single}
V.~Chinni, P.~Latzel, M.~Z{\'e}gaoui, C.~Coinon, X.~Wallart, E.~Peytavit,
  J.~Lampin, K.~Engenhardt, P.~Szriftgiser, M.~Zaknoune \emph{et~al.},
  \enquote{{Single-channel 100 Gbit/s transmission using III--V UTC-PDs for
  future IEEE 802.15. 3d wireless links in the 300 GHz band},}
  {\protect\JournalTitle{Electronics Letters}} \textbf{54}, 638--640 (2018).

\bibitem{hamada2018300}
H.~Hamada, T.~Fujimura, I.~Abdo, K.~Okada, H.-J. Song, H.~Sugiyama,
  H.~Matsuzaki, and H.~Nosaka, \enquote{300-{GHz. 100-Gb/s InP-HEMT wireless
  transceiver using a 300-GHz} fundamental mixer,} in \emph{2018 IEEE/MTT-S
  International Microwave Symposium-IMS,}  (IEEE, 2018), pp. 1480--1483.

\bibitem{Chen2018Influence}
J.~Chen, D.~Kuylenstierna, S.~E. Gunnarsson, Z.~S. He, T.~Eriksson, T.~Swahn,
  and H.~Zirath, \enquote{Influence of {White LO Noise on Wideband
  Communication},} {\protect\JournalTitle{IEEE Transactions on Microwave Theory
  and Techniques}} \textbf{66}, 3349--3359 (2018).

\bibitem{nagatsuma2013terahertz}
T.~Nagatsuma, S.~Horiguchi, Y.~Minamikata, Y.~Yoshimizu, S.~Hisatake,
  S.~Kuwano, N.~Yoshimoto, J.~Terada, and H.~Takahashi, \enquote{Terahertz
  wireless communications based on photonics technologies,}
  {\protect\JournalTitle{Optics Express}} \textbf{21}, 23736--23747 (2013).

\bibitem{harter2019wireless}
T.~Harter, S.~Ummethala, M.~Blaicher, S.~Muehlbrandt, S.~Wolf, M.~Weber,
  M.~M.~H. Adib, J.~N. Kemal, M.~Merboldt, F.~Boes \emph{et~al.},
  \enquote{Wireless {THz} link with optoelectronic transmitter and receiver,}
  {\protect\JournalTitle{Optica}} \textbf{6}, 1063--1070 (2019).

\bibitem{li2019low}
Y.~Li, A.~Rolland, K.~Iwamoto, N.~Kuse, M.~Fermann, and T.~Nagatsuma,
  \enquote{Low-noise millimeter-wave synthesis from a dual-wavelength fiber
  {Brillouin} cavity,} {\protect\JournalTitle{Optics Letters}} \textbf{44},
  359--362 (2019).

\bibitem{herr2014temporal}
T.~Herr, V.~Brasch, J.~D. Jost, C.~Y. Wang, N.~M. Kondratiev, M.~L. Gorodetsky,
  and T.~J. Kippenberg, \enquote{Temporal solitons in optical microresonators,}
  {\protect\JournalTitle{Nature Photonics}} \textbf{8}, 145--152 (2014).

\bibitem{brasch2016photonic}
V.~Brasch, M.~Geiselmann, T.~Herr, G.~Lihachev, M.~H. Pfeiffer, M.~L.
  Gorodetsky, and T.~J. Kippenberg, \enquote{Photonic chip--based optical
  frequency comb using soliton {Cherenkov} radiation,}
  {\protect\JournalTitle{Science}} \textbf{351}, 357--360 (2016).

\bibitem{zhang2019terahertz}
S.~Zhang, J.~M. Silver, X.~Shang, L.~Del~Bino, N.~M. Ridler, and P.~Del’Haye,
  \enquote{Terahertz wave generation using a soliton microcomb,}
  {\protect\JournalTitle{Optics Express}} \textbf{27}, 35257--35266 (2019).

\bibitem{huang2017globally}
S.-W. Huang, J.~Yang, S.-H. Yang, M.~Yu, D.-L. Kwong, T.~Zelevinsky,
  M.~Jarrahi, and C.~W. Wong, \enquote{Globally stable microresonator {Turing}
  pattern formation for coherent high-power {THz} radiation on-chip,}
  {\protect\JournalTitle{Physical Review X}} \textbf{7}, 041002 (2017).

\bibitem{tetsumoto2020300}
T.~Tetsumoto, F.~Ayano, M.~Yeo, J.~Webber, T.~Nagatsuma, and A.~Rolland,
  \enquote{300 {GHz} wave generation based on a kerr microresonator frequency
  comb stabilized to a low noise microwave reference,}
  {\protect\JournalTitle{Optics Letters}} \textbf{45}, 4377--4380 (2020).

\bibitem{wang2021towards}
B.~Wang, J.~S. Morgan, K.~Sun, M.~Jahanbozorgi, Z.~Yang, M.~Woodson,
  S.~Estrella, A.~Beling, and X.~Yi, \enquote{Towards high-power,
  high-coherence, integrated photonic mmwave platform with microcavity
  solitons,} {\protect\JournalTitle{Light: Science \& Applications}}
  \textbf{10}, 1--10 (2021).

\bibitem{tetsumoto2021optically}
T.~Tetsumoto, T.~Nagatsuma, M.~E. Fermann, G.~Navickaite, M.~Geiselmann, and
  A.~Rolland, \enquote{Optically referenced 300~{GHz} millimetre-wave
  oscillator,} {\protect\JournalTitle{Nature Photonics}} \textbf{15}, 516--522
  (2021).

\bibitem{kuo2010spectral}
F.-M. Kuo, J.-W. Shi, H.-C. Chiang, H.-P. Chuang, H.-K. Chiou, C.-L. Pan, N.-W.
  Chen, H.-J. Tsai, and C.-B. Huang, \enquote{Spectral power enhancement in a
  100 {GHz} photonic millimeter-wave generator enabled by spectral line-by-line
  pulse shaping,} {\protect\JournalTitle{IEEE Photonics Journal}} \textbf{2},
  719--727 (2010).

\bibitem{kuse2022low}
N.~Kuse, K.~Nishimoto, Y.~Tokizane, S.~Okada, K.~Minoshima, and T.~Yasui,
  \enquote{Low noise 560 {GHz} generation from a fiber-referenced kerr
  microresonator soliton comb,} in \emph{CLEO: Applications and Technology,}
  (Optica Publishing Group, 2022), pp. JW3B--1.

\bibitem{stone2020harnessing}
J.~R. Stone and S.~B. Papp, \enquote{Harnessing dispersion in soliton
  microcombs to mitigate thermal noise,} {\protect\JournalTitle{Physical Review
  Letters}} \textbf{125}, 153901 (2020).

\bibitem{yi2017single}
X.~Yi, Q.-F. Yang, X.~Zhang, K.~Y. Yang, X.~Li, and K.~Vahala,
  \enquote{Single-mode dispersive waves and soliton microcomb dynamics,}
  {\protect\JournalTitle{Nature Communications}} \textbf{8}, 1--9 (2017).

\bibitem{tetsumoto2021effects}
T.~Tetsumoto, J.~Jiang, M.~E. Fermann, G.~Navickaite, M.~Geiselmann, and
  A.~Rolland, \enquote{Effects of a quiet point on a {Kerr} microresonator
  frequency comb,} {\protect\JournalTitle{OSA Continuum}} \textbf{4},
  1348--1357 (2021).

\bibitem{urick2015fundamentals}
V.~J. Urick, K.~J. Williams, and J.~D. McKinney, \emph{{Fundamentals of
  Microwave Photonics}} (John Wiley \& Sons, 2015).

\bibitem{briles2018interlocking}
T.~C. Briles, J.~R. Stone, T.~E. Drake, D.~T. Spencer, C.~Fredrick, Q.~Li,
  D.~Westly, B.~Ilic, K.~Srinivasan, S.~A. Diddams \emph{et~al.},
  \enquote{Interlocking kerr-microresonator frequency combs for microwave to
  optical synthesis,} {\protect\JournalTitle{Optics Letters}} \textbf{43},
  2933--2936 (2018).

\bibitem{kuse2019control}
N.~Kuse, T.~C. Briles, S.~B. Papp, and M.~E. Fermann, \enquote{{Control of
  Kerr-microresonator optical frequency comb by a dual-parallel Mach-Zehnder
  interferometer},} {\protect\JournalTitle{Optics Express}} \textbf{27},
  3873--3883 (2019).

\bibitem{shafik2006extended}
R.~A. Shafik, M.~S. Rahman, and A.~R. Islam, \enquote{On the extended
  relationships among {EVM}, {BER} and {SNR} as performance metrics,} in
  \emph{2006 International Conference on Electrical and Computer Engineering,}
  (IEEE, 2006), pp. 408--411.

\bibitem{chang2010forward}
F.~Chang, K.~Onohara, and T.~Mizuochi, \enquote{Forward error correction for
  100 {G} transport networks,} {\protect\JournalTitle{IEEE Communications
  Magazine}} \textbf{48}, S48--S55 (2010).

\bibitem{keysight_M8196A}
\enquote{{Keysight Technologies M8196A},}
  \url{https://www.keysight.com/us/en/assets/7018-04911/data-sheets/5992-0971.pdf}.

\bibitem{dan2020superheterodyne}
I.~Dan, G.~Ducournau, S.~Hisatake, P.~Szriftgiser, R.-P. Braun, and
  I.~Kallfass, \enquote{A superheterodyne 300 {GHz} wireless link for
  ultra-fast terahertz communication systems,}
  {\protect\JournalTitle{International Journal of Microwave and Wireless
  Technologies}} \textbf{12}, 578--587 (2020).

\bibitem{liang2021recent}
D.~Liang and J.~E. Bowers, \enquote{Recent {Progress in Heterogeneous
  III-V-on-Silicon Photonic Integration},} {\protect\JournalTitle{Light:
  Advanced Manufacturing}} \textbf{2}, 1--25 (2021).

\bibitem{kaur2021hybrid}
P.~Kaur, A.~Boes, G.~Ren, T.~G. Nguyen, G.~Roelkens, and A.~Mitchell,
  \enquote{Hybrid and heterogeneous photonic integration,}
  {\protect\JournalTitle{APL Photonics}} \textbf{6}, 061102 (2021).

\bibitem{xiang2021laser}
C.~Xiang, J.~Liu, J.~Guo, L.~Chang, R.~N. Wang, W.~Weng, J.~Peters, W.~Xie,
  Z.~Zhang, J.~Riemensberger \emph{et~al.}, \enquote{Laser soliton microcombs
  heterogeneously integrated on silicon,} {\protect\JournalTitle{Science}}
  \textbf{373}, 99--103 (2021).

\bibitem{spencer2018optical}
D.~T. Spencer, T.~Drake, T.~C. Briles, J.~Stone, L.~C. Sinclair, C.~Fredrick,
  Q.~Li, D.~Westly, B.~R. Ilic, A.~Bluestone \emph{et~al.}, \enquote{An
  optical-frequency synthesizer using integrated photonics,}
  {\protect\JournalTitle{Nature}} \textbf{557}, 81--85 (2018).

\bibitem{liu2020photonic}
J.~Liu, E.~Lucas, A.~S. Raja, J.~He, J.~Riemensberger, R.~N. Wang, M.~Karpov,
  H.~Guo, R.~Bouchand, and T.~J. Kippenberg, \enquote{Photonic microwave
  generation in the {X-and K-band} using integrated soliton microcombs,}
  {\protect\JournalTitle{Nature Photonics}} \textbf{14}, 486--491 (2020).

\bibitem{zhang2021integrated}
M.~Zhang, C.~Wang, P.~Kharel, D.~Zhu, and M.~Lon{\v{c}}ar, \enquote{Integrated
  lithium niobate electro-optic modulators: when performance meets
  scalability,} {\protect\JournalTitle{Optica}} \textbf{8}, 652--667 (2021).

\bibitem{stern2018battery}
B.~Stern, X.~Ji, Y.~Okawachi, A.~L. Gaeta, and M.~Lipson,
  \enquote{Battery-operated integrated frequency comb generator,}
  {\protect\JournalTitle{Nature}} \textbf{562}, 401--405 (2018).

\bibitem{jang2021conversion}
J.~K. Jang, Y.~Okawachi, Y.~Zhao, X.~Ji, C.~Joshi, M.~Lipson, and A.~L. Gaeta,
  \enquote{Conversion efficiency of soliton kerr combs,}
  {\protect\JournalTitle{Optics Letters}} \textbf{46}, 3657--3660 (2021).

\bibitem{xue2017microresonator}
X.~Xue, P.-H. Wang, Y.~Xuan, M.~Qi, and A.~M. Weiner, \enquote{Microresonator
  {Kerr} frequency combs with high conversion efficiency,}
  {\protect\JournalTitle{Laser \& Photonics Reviews}} \textbf{11}, 1600276
  (2017).

\bibitem{kim2019turn}
B.~Y. Kim, Y.~Okawachi, J.~K. Jang, M.~Yu, X.~Ji, Y.~Zhao, C.~Joshi, M.~Lipson,
  and A.~L. Gaeta, \enquote{Turn-key, high-efficiency {Kerr} comb source,}
  {\protect\JournalTitle{Optics Letters}} \textbf{44}, 4475--4478 (2019).

\bibitem{xue2019super}
X.~Xue, X.~Zheng, and B.~Zhou, \enquote{Super-efficient temporal solitons in
  mutually coupled optical cavities,} {\protect\JournalTitle{Nature Photonics}}
  \textbf{13}, 616--622 (2019).

\bibitem{helgason2022power}
{\'O}.~B. Helgason, M.~Girardi, Z.~Ye, J.~Schr{\"o}der \emph{et~al.},
  \enquote{Power-efficient soliton microcombs in anomalous-dispersion photonic
  molecules,} in \emph{CLEO: QELS\_Fundamental Science,}  (Optica Publishing
  Group, 2022), pp. FW4J--5.

\bibitem{matlab_example}
\enquote{{QPSK Transmitter and Receiver in Simulink},}
  \url{https://www.mathworks.com/help/supportpkg/usrpradio/ug/qpsk-transmitter-with-usrp-r-hardware-1.html}.

\bibitem{matlab_example2}
\enquote{{QPSK Receiver with USRP\textregistered ~Hardware},}
  \url{https://www.mathworks.com/help/supportpkg/usrpradio/ug/qpsk-receiver-with-usrp-r-hardware-1.html}.

\end{thebibliography}

\begin{backmatter}
\bmsection{Acknowledgments}
We appreciate Professor Nagatsuma for loan of a 300~GHz Schottky diode.

\bmsection{Disclosures}
The authors declare no conflicts of interest.

\bmsection{Data availability} Data underlying the results presented in this paper are not publicly available at this time but may be obtained from the authors upon reasonable request.

\newpage

\renewcommand{\thefigure}{s\arabic{figure}}
\renewcommand{\thesection}{S\arabic{section}}
\setcounter{figure}{0}
\setcounter{section}{0}

\title{Supplementary material for ``A 300~GHz wireless link based on an integrated Kerr soliton comb''}

\author{Tomohiro Tetsumoto,\authormark{1} and Antoine Rolland\authormark{1*}}

\address{\authormark{1}IMRA America Inc., Boulder Research Labs, 1551 South Sunset St, Suite C, Longmont, Colorado 80501, USA}

\email{\authormark{*}arolland@imra.com} 

\section{Phase noise characterization}
Phase noise measurement is performed by using the setup in Fig.~2(a) in the main text and in Fig.~\ref{supp_fig1}(a). 
First, the beat signal, given by Eq.~6 in the main text, is evaluated in the unlocked and locked states.
The unlocked signal gives the free-running microcomb noise (black curve) as far as the calibrated 10~GHz SYN noise to 300~GHz (green curve) is lower than it.
The two curves meet at around 100~kHz, and the SYN noise dominates the free-running noise above that.
When the system is locked, the beat signal gives the in-loop noise (red curve). 
We tuned the locking bandwidth to around the cross point of the free-running and the 10~GHz SYN noise. 
Then, we performed a measurement in the 300~GHz wave domain to evaluate the stabilized microcomb noise with Fig.~\ref{supp_fig1}(a).
Two 300~GHz waves, with a slight frequency difference, are generated through photo-mixing the stabilized microcomb and two lines of an EO comb prepared for this measurement at two UTC-PDs, respectively.
They are mixed, and the phase noise of the mixer output is characterized (blue curve).
The measurement was performed twice using two oscillators as the EO comb's drivers. 
The two results are stitched to maximize the sensitivity (the orange curve gives the sensitivity).
The phase noise of the mixer output follows the calibrated 10~GHz SYN noise from 100~Hz to about 400~Hz until it is limited by the local oscillator noise for the measurement between 400~Hz and 20~kHz.
From 20~kHz to 100~kHz, it is limited by the in-loop noise.
Above the offset frequency of 100~kHz (out of the loop bandwidth), the mixer output presents a free-running microcomb noise, which is not measured in the black line limited by the sensitivity in the measurement (green curve).
In addition, we evaluate the frequency instability of the mixer output, which is shown in Fig.~\ref{supp_fig1}(c) (red curve), along with the frequency instability plots shown in the main text.
Note that the EO comb driver is synchronized to the same clock as the microcomb.
It shows more than 100 times lower stability than the free-running and is averaged down over the measurement time.
This indicates that the two oscillators are synchronized well, as expected.

\begin{figure*}[!ht]
    \centering
    \includegraphics[width=\linewidth]{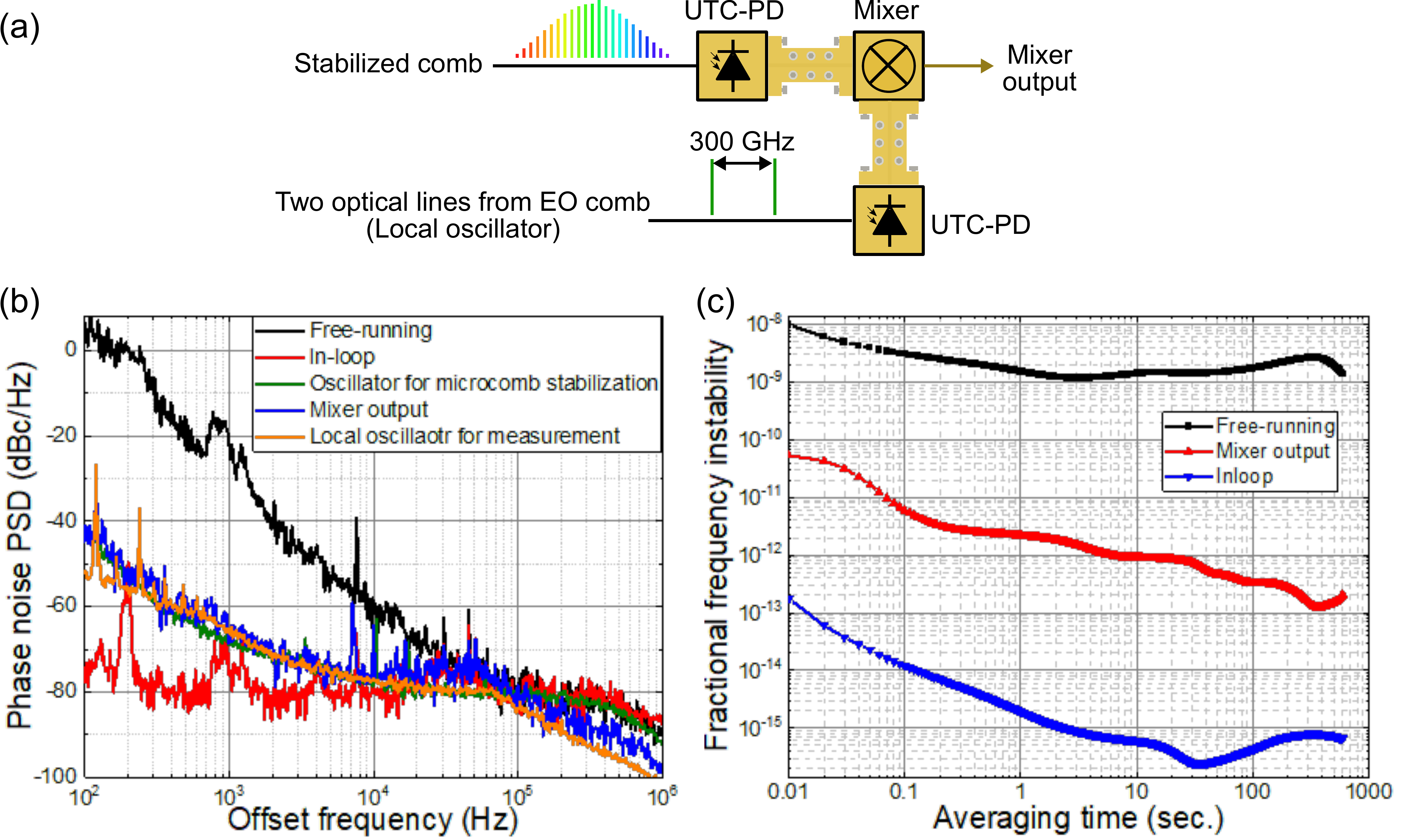}
    \caption{\footnotesize{(\textbf{a}) Schematic drawing of an experimental setup for phase noise measurement.  (\textbf{b}) Measured phase noise. (\textbf{c}) Measured frequency instability.}}
    \label{supp_fig1}
\end{figure*} 

\section{Program for digital signal processing}
The DSP in this study is basically based on example models in MATLAB \& Simulink \cite{matlab_example, matlab_example2} with some modifications. Figure~\ref{supp_fig2} is the block diagram of the functions.
The upper row is for the transmitter, and the lower is for the receiver. In the transmitter, Bernoulli binary series, based on \textit{rand()} function in MATLAB, is generated and encoded into n-QAM symbols (down-sampled by a factor of log2(n)).
The symbol rate is 200~kHz.
Then, a raised cosine transmit filter with a roll-off factor of 0.5 is applied to it, where the signal is up-sampled by a factor of log2(n).
The data is sent to the FPGA transmitter, where the transmitter outputs the produced modulation signal to the IM.
The amplitude of received signals is stabilized via auto gain control (AGC block), whose output is sent to a raised cosine transmit filter with a roll-off factor of 0.5.
The frequency offset of the signal is estimated with the "Coarse Frequency Compensation" block, and the output is sent to the blocks for symbol timing, and carrier frequency and phase synchronization.
The loop bandwidth for the synchronization processes, normalized by the sample rate, is 0.001.
Finally, the imbalance between in-phase and quadrature components is corrected manually, and the constellation diagram is displayed with calculated EVM.

\begin{figure*}[!ht]
    \centering
    \includegraphics[width=\linewidth]{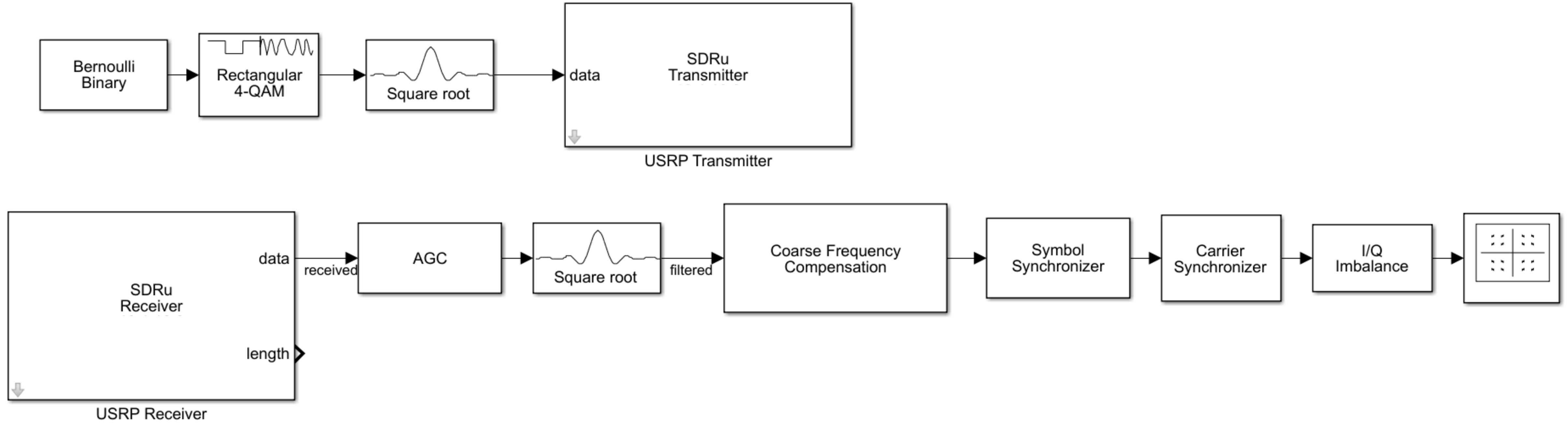}
    \caption{\footnotesize{Block diagram for DSP.}}
    \label{supp_fig2}
\end{figure*}

\end{backmatter}

\end{document}